\documentclass[journal,10pt]{IEEEtran}

\ifCLASSINFOpdf
\else
   \usepackage[dvips]{graphicx}
\fi
\usepackage{url}
\usepackage{amssymb}
\usepackage{bbding}
\usepackage{graphicx}
\usepackage{lettrine}
\usepackage[numbers,sort&compress]{natbib}
\usepackage{amsmath} 
\usepackage{amssymb} 
\usepackage{amsfonts}
\usepackage{booktabs}
\usepackage{hyperref}
\usepackage{footmisc}
\begin{document}

\title{Adaptive Mixture of Low-Rank Experts \\for Robust Audio Spoofing Detection}

\author{Qixian Chen$^{\dagger}$, Yuxiong Xu$^{\dagger}$, Sara Mandelli, \IEEEmembership{Member, IEEE}, Sheng Li, and Bin Li\Envelope, \IEEEmembership{Senior Member, IEEE}
\thanks{Qixian Chen, Yuxiong Xu, and Bin Li are with the Shenzhen Key Laboratory of Media Security, College of Electronics and Information Engineering, Shenzhen University, Shenzhen 518060, China (emails: \href{mailto:2022280087@email.szu.edu.cn}{2022280087@email.szu.edu.cn}, \href{mailto:xuyuxiong2022@email.szu.edu.cn}{xuyuxiong2022@email.szu.edu.cn}, \href{mailto:libin@szu.edu.cn}{libin@szu.edu.cn}). Sara Mandelli is with the DEIB, Politecnico di Milano, Milan, Italy (email: \href{mailto:sara.mandelli@polimi.it}{sara.mandelli@polimi.it}). Sheng Li is with the Afirstsoft Technology Group Co., Ltd, Shenzhen, China (email: \href{mailto:admin@tenorshare.cn}{admin@tenorshare.cn}).}
\thanks{\textsuperscript{$\dagger$} Equal contributions. \textsuperscript{\Envelope} Corresponding author.}
}

\markboth{Journal of \LaTeX\ Class Files, Vol. 14, No. 8, August 2015}
{Shell \MakeLowercase{\textit{et al.}}: Bare Demo of IEEEtran.cls for IEEE Journals}
\maketitle

\begin{abstract}
In audio spoofing detection, most studies rely on clean datasets, making models susceptible to real-world post-processing attacks, such as channel compression and noise. To overcome this challenge, we propose the Adaptive MixtUre Low-rank ExperTs (AMULET) framework, which enhances resilience by leveraging attack-specific knowledge and dynamically adapting to varied attack conditions. Specifically, AMULET employs Attack-Specific Experts (ASEs) fine-tuned with Low-Rank Adaptation (LoRA), allowing each expert to focus on distinct post-processing patterns using just 1.13\% of the parameters required for full fine-tuning. Furthermore, we introduce Adaptive Expert Fusion (AEF), which adaptively selects and integrates expert knowledge to enhance the robustness of spoofing detection. Experimental results demonstrate that AMULET significantly enhances robustness by improving noise resilience and exhibiting greater adaptability to unseen post-processing methods compared to models  trained with full fine-tuning. Additionally, our framework outperforms both single expert and other expert aggregation strategies under various mixed attacks, demonstrating its superior robustness and adaptability in managing complex real-world scenarios.
\end{abstract}

\begin{IEEEkeywords}
 audio spoofing detection, attack-specific expert, mixture of experts, adaptive expert fusion 
\end{IEEEkeywords}

\IEEEpeerreviewmaketitle

\section{Introduction}
\lettrine[lines=2]{\textbf{R}}{ECENT} advancements in generative artificial intelligence have significantly transformed speech synthesis and manipulation, offering new opportunities while simultaneously posing security risks to Automatic Speaker Verification (ASV) systems. The primary challenge lies in developing robust audio spoofing detection methods that can effectively counter evolving attacks \cite{Xu2024}.
Moreover, real-world factors such as noise, codec artifacts, and channel distortions further complicate the detection process. Given the complex nature of spoofing variations, it is essential to develop detection methods capable of performing well, even with limited training data.

While substantial progress has been made in ASVspoof detection,
current approaches still exhibit several critical limitations: (i) excessive reliance on clean laboratory datasets \cite{wang22_odyssey}, (ii) insufficient adaptability to distributional shifts in real-world scenarios \cite{mittal2022static}, and (iii) limited robustness against post-processing distortions such as noise and compression \cite{li2024audio}. 
Detection models trained on clean audio often exhibit limited robustness when exposed to real-world degradations, prompting efforts to develop more resilient methods.

To address noise-related distortions, speech enhancement techniques \cite{zhang2022time} are frequently adopted as preprocessing steps to recover cleaner spectrograms from noisy speech signals \cite{cai2020within, shon2019voiceid}. 
However, prior studies have shown that such techniques can introduce estimation errors and undesired artifacts \cite{fan2022specmnet, fan2023two}, which, in some cases, are more harmful than the original noise itself \cite{Iwamoto2022HowBA}. 
Tak et al. \cite{tak2022automatic} improved models robustness
by leveraging the pre-trained Wav2Vec 2.0 model and applying data augmentation techniques. 
Nevertheless, their approach primarily focuses on communication-induced degradations, with limited consideration of other post-processing operations such as compression and filtering.
As a result, existing enhancement-based methods tend to be narrowly targeted and remain insufficient in addressing the broad spectrum of post-processing attacks encountered in practical scenarios.

In this context, the Mixture of Experts (MoE) framework \cite{jacobs1991adaptive}, which consists of multiple specialized sub-models (i.e., experts), has attracted increasing attention in deepfake detection. Recent studies have leveraged MoE to improve generalization across diverse domains \cite{dai2021generalizable, xu2022mimic}. For example, Zhou et al. \cite{zhou2022adaptive} proposed an adaptive MoE learning strategy that effectively incorporates domain-specific knowledge to bridge the gap between seen and unseen attack domains. Wang et al. \cite{wang2025mixture} introduced a feature fusion approach that aggregates layer-wise representations from a pretrained model to enhance detection. Negroni et al. \cite{negroni2025leveraging} extended the MoE framework to speech deepfake detection, demonstrating the benefits of expert specialization in handling diverse spoofing conditions.

Although MoE architectures have shown promising results in deepfake detection, the efficient use of large-scale expert models remains insufficiently explored. 
To address this gap, we propose the Adaptive MixtUre Low-rank ExperTs (AMULET) framework, which leverages Low-Rank Adaptation (LoRA) \cite{hu2022lora} to enable robust and efficient audio spoofing detection using large expert models.
AMULET combines a fully fine-tuned shared expert together with lightweight Attack-Specific Experts (ASEs), which are adapted via LoRA, enabling to limit the fine-tuning overhead.
Furthermore, we propose an Adaptive Expert Fusion (AEF) module which utilizes a simple linear gating network to dynamically select the most relevant experts, avoiding the computational burden associated with attention-based mechanisms. This architecture improves robustness against a wide range of spoofing attacks and ensures stable performance under various post-processing distortions.

Our contribution is summarized below:
\begin{itemize}
    \item We propose AMULET, a novel framework for robust audio spoofing detection that incorporates domain-specific knowledge through ASEs, effectively bridging the gap between seen and unseen post-processing attacks.
    \item 
    We utilize LoRA to fine-tune ASEs efficiently, enabling each expert to capture distinct post-processing characteristics with minimal computational overhead.
    \item 
    We introduce AEF, a dynamic fusion mechanism that selectively integrates expert knowledge to enhance generalization to unseen attack types, significantly improving robustness under real-world post-processing conditions.
\end{itemize}

\section{Methodology}

\begin{figure}[t]
  \centering 
  \includegraphics[width=0.45\textwidth]{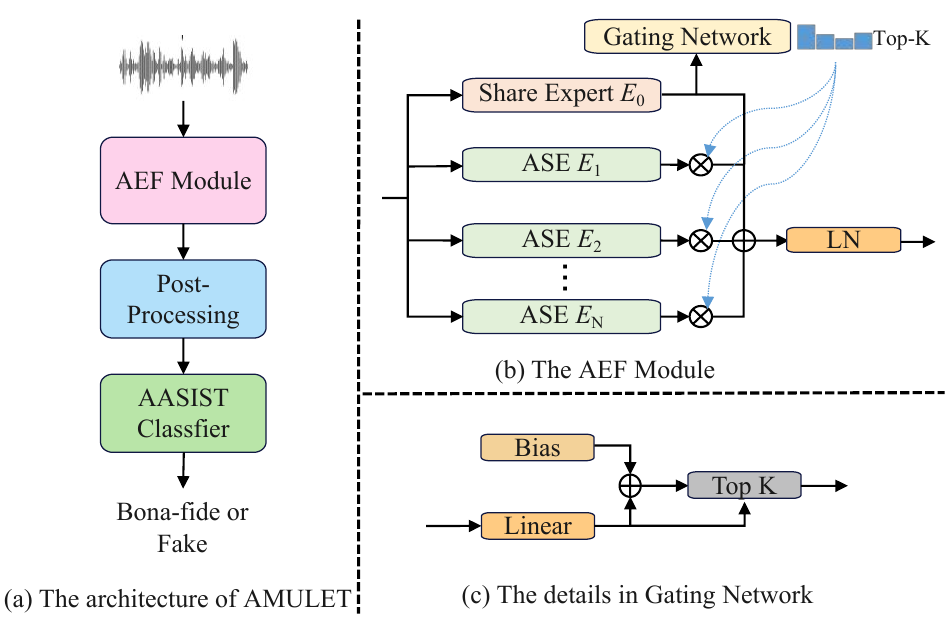}
  \caption{Overview of the AMULET pipeline. }
    \vspace{-15pt}
  \label{fig:MoE} 
\end{figure}

\subsection{Overview of Adaptive Mixture of Low-Rank Experts}

The AMULET framework enhances robustness in audio spoofing detection through two key procedures: training Attack-Specific Experts (ASEs) and employing Adaptive Expert Fusion (AEF). The pipeline is shown in Fig.~\ref{fig:MoE}. 

During ASEs training, a shared expert is first fine-tuned on clean data and then adapted to specific attack types using LoRA. 
Unlike \cite{zhou2022adaptive}, where computational cost grows linearly with expert model size, our method avoids such overhead, ensuring efficient adaptation to various attacks.

In the AEF module, 
instead of using parameter-free attention weights as in~\cite{zhou2022adaptive}, we employ a gating network that dynamically selects the most relevant ASEs based on input characteristics.
The outputs of the selected experts, combined with shared features, are passed through the same post-processing used in~\cite{tak2022automatic} to match the input feature size required by the classification head. The resulting features are then fed into the classification head for final detection.

\vspace{-5pt}

\subsection{Shared Expert and  Attack-Specific Experts (ASEs)}
 We adopt the expert framework model described in \cite{tak2022automatic, DBLP:journals/corr/abs-2408-09933}, which integrates Wav2Vec 2.0 (W2V2) \cite{baevski2020wav2vec} 
 as the front-end feature extractor and graph attention networks (i.e., AASIST) \cite{jung2022aasist} as the back-end classifier.

In the initial training phase, we fully fine-tune the entire detection model to obtain a shared expert \( E_0 \) which captures ``clean'' spoofing detection features. To do so, we use 
the ASVspoof 2019 LA training and development datasets~\cite{wang2020asvspoof}.
Specifically, we adjust all parameters of the model \( \boldsymbol{\theta} \), which includes both the front-end extractor and back-end classifier. We define as $\boldsymbol{\theta}^*$ the fully fine-tuned shared expert parameters.


Then, we train $N$ Attack-Specific Experts (ASEs) considering specific attacks applied to the same dataset. The attacks include several audio perturbations, which we describe in detail in Section~\ref{subsec:experiments_dataset}. 
Existing MoE-based audio spoofing detection models require full fine-tuning of the ASE networks \cite{wang2025mixture, negroni2025leveraging}, leading to significant computational overhead and scalability challenges as the size of expert models increases. To address this issue, 
we adopt a LoRA-based fine-tuning strategy \cite{hu2022lora} to train our ASEs. For each ASE \( E_i \), instead of fine-tuning all parameters, we introduce low-rank matrices to modify only the linear layer parameters of the W2V2 model. 

Specifically, we define the low-rank matrices as \( \mathbf{A}_i \in \mathbb{R}^{m \times r} \) and \( \mathbf{B}_i \in \mathbb{R}^{r \times n} \), where $r$ is the rank, with \( m \) and \( n \) corresponding to the input and output dimensions of the W2V2 model's linear layer.  
The original linear layer parameters of the W2V2 model are denoted as \( \boldsymbol{\boldsymbol{\theta}}^*_0 \), and they are modified by LoRA as
\begin{equation}
\boldsymbol{\theta}_i = \boldsymbol{\theta}^*_ 0+ \alpha \times (\mathbf{A}_i \mathbf{B}_i) , \quad i \in \{1, \dots, N\}, 
\end{equation}
where $\alpha$ is a scaling factor, and 
\( \mathbf{A}_i \mathbf{B}_i \) denotes standard matrix multiplication.
The resulting fine-tuned linear layer parameters are denoted by \( \boldsymbol{\theta}_i \).

\vspace{-5pt}
\subsection{Adaptive Expert Fusion (AEF)}


Inspired by DeepSeek V3 \cite{liu2024deepseek}, we propose the Adaptive Expert Fusion (AEF) framework, which dynamically manages the activation of ASEs based on input characteristics during both training and inference. Under AEF, the shared expert \( E_0 \) is always active to provide generalizable features, while the ASEs \( \{E_i\} \) are selectively activated according to the input.

All experts, with their parameters frozen, process the same input audio \( \mathbf{x} \), generating feature representations from the final layer of the W2V2 model, with size \( b \times 201 \times 1024 \), where \( b \) is the batch size, \( 201 \) is the number of time steps, and \( 1024 \) is the feature dimension per time step:
\begin{equation}
\mathbf{z}_i = E_i(\mathbf{x}), \quad i \in \{0, 1, \dots, N\}.
\end{equation}

The gating network \( G \) computes the experts contribution weights from the shared expert output, in order to incorporate the features from the ASEs which are missed in shared expert:
\begin{equation}
\mathbf{w} = G(\mathbf{z}_0) = G(E_0(\mathbf{x})), \quad \mathbf{w} \in \mathbb{R}^N.
\end{equation}

Here, \( \mathbf{w} = [\mathbf{w}_1, \mathbf{w}_2, \dots, \mathbf{w}_N] \) denotes the vector of contribution weights for all ASEs, where each \( \mathbf{w}_i \) indicates the contribution of expert \( E_i \).

\begin{table*}[t]
\centering
\caption{Performance comparison of experts, expert aggregation strategies, and our method in terms of EER (\%). }
\vspace{-5pt}
\label{tab:Mode Performance}
\footnotesize
\begin{tabular}{lcccccccc} 
\toprule
Model          & T0             & T1             & T2             & T3             & T4             & T5             & T6             & Avg.             \\ 
\midrule
\( E_0 \)  
& 0.26~          & 7.63~          & 11.73~         & 0.59~          & 1.13~          & 4.79~          & 5.24~          & 4.48~           \\
\( E_1 \) 
& 0.33~          & 0.52~          & 0.85~          & 0.45~          & 1.17~          & 2.26~          & 1.21~          & 0.97~           \\
\( E_2 \) 
& 0.28~          & 2.19~          & \underline{0.26}~  & 0.33~          & 1.06~          & 2.82~          & 1.31~          & 1.18~           \\
\( E_3 \) 
& 0.29~          & 5.06~          & 8.04~          & \underline{0.30}~  & 0.98~          & 3.99~          & 1.96~          & 2.95~           \\
\( E_4 \) 
& \textbf{0.18~} & 6.81~          & 14.98~         & 0.35~          & \textbf{0.35}~ & 5.48~          & 5.59~          & 4.82~           \\
\( E_5 \) 
& \underline{0.23}~  & 1.43~          & 4.57~          & \textbf{0.25}~ & 0.72~          & \textbf{0.58}~ & 1.38~          & 1.31~           \\ \midrule
\textit{Ensemble}       & \textbf{0.18~} & 1.43~          & \textbf{0.23~} & 0.37~          & \underline{0.41}~  & 2.75~          & 1.48~          & 0.98~           \\
\textit{Attention} \cite{zhou2022adaptive}& 0.53~          & 1.00~          & 0.70~          & 0.58~          & 0.86~          & 2.57~          & 1.20~          & 1.06~           \\
\textit{MoEF} \cite{wang2025mixture}& 0.45~          & 1.98~          & 1.46~          & 0.35~          & 0.47~          & 2.30~          & 1.46~          & 1.21~           \\ \midrule
\( E_0 \) + top-3 (Ours) & 0.44~          & 0.57~          & 0.27~          & 0.38~          & 0.68~          & 0.94~          & 1.19~          & 0.64~           \\
\( E_0 \) + top-4 (Ours) & 0.29~          & \underline{0.49~}  & 0.29~          & 0.31~          & 0.81~          & 1.08~          & \underline{1.17}~  & \underline{0.63}~   \\
\( E_0 \) + top-5 (Ours) & 0.27~          & \textbf{0.43~} & \underline{0.26}~  & \underline{0.30}~  & 0.65~          & \underline{0.83}~  & \textbf{1.10~} & \textbf{0.55}~  \\
\bottomrule
\end{tabular}
\vspace{-12pt}
\end{table*}

\begin{table}[t]
\centering
\caption{EER (\%) achieved by the experts: FFT vs. LoRA. TP denotes the number of trainable parameters.}
\vspace{-5pt}
\label{tab:experts performance}
\footnotesize
\begin{tabular}{lcccccc} 
\toprule
Model     & TP & T1            & T2            & T3            & T4            & T5             \\ \midrule
\( E_1 \) (FFT)   & 318M         & 0.59          & 7.43          & 1.85          & 6.17          & 4.78           \\
\( E_1 \) (LoRA) & 3.59M           & \textbf{0.52} & \textbf{0.85} & \textbf{0.45} & \textbf{1.17} & \textbf{2.26}  \\ \midrule
\( E_2 \) (FFT)   & 318M         & \textbf{1.71} & \textbf{0.26} & 0.55          & 1.61          & 4.93           \\
\( E_2 \) (LoRA) & 3.59M           & 2.19          & \textbf{0.26} & \textbf{0.33} & \textbf{1.06} & \textbf{2.82}  \\ \midrule
\( E_3 \) (FFT)   & 318M         & \textbf{2.78} & \textbf{2.72} & \textbf{0.14} & \textbf{0.19} & 5.83           \\
\( E_3 \) (LoRA) & 3.59M           & 5.06          & 8.04          & 0.30          & 0.98          & \textbf{3.99}  \\ \midrule
\( E_4 \) (FFT)   & 318M         & \textbf{2.45} & \textbf{2.46} & \textbf{0.18} & \textbf{0.18} & 5.63           \\
\( E_4 \) (LoRA) & 3.59M           & 6.81          & 14.98         & 0.35          & 0.35          & \textbf{5.48}  \\ \midrule
\( E_5 \) (FFT)   & 318M         & 1.62          & 7.36          & 1.15          & 2.62          & 1.13           \\
\( E_5 \) (LoRA) & 3.59M           & \textbf{1.43} & \textbf{4.57} & \textbf{0.25} & \textbf{0.72} & \textbf{0.58}  \\
\bottomrule
\end{tabular}
\vspace{-3pt}
\end{table}

To enhance efficiency and mitigate overfitting, we apply a top-$k$ expert selection strategy, activating only the \( k\) experts with the highest gating scores. Let \( \mathcal{T}_k \) denote the indices of the selected experts. The final aggregated representation is:
\begin{equation}
\mathbf{z}_{\text{MoE}} = LN \left( \sum_{i \in \mathcal{T}_k} \mathbf{w}_i \mathbf{z}_i + \mathbf{z}_0 \right),
\end{equation}
where \( \mathbf{z}_i \) and \( \mathbf{z}_0 \) represent the feature representations of the ASEs and the shared expert, respectively. 
\( LN \) refers to Layer Normalization \cite{ba2016layer}, which normalizes the input features to zero mean and unit variance along the channel dimension. 

The fused vector \( \mathbf{z}_{\text{MoE}} \) is then processed by a post-processing operator $P$ to match the input feature size required by the classification head \(C\), which subsequently produces the final prediction:
\begin{equation}
y = C(P(\mathbf{z}_{\text{MoE}})).
\end{equation}

Both \( P \) and \( C \) adopt the same design as described in~\cite{tak2022automatic}.

\section{Experiments}
\subsection{Datasets}
\label{subsec:experiments_dataset}
For training the shared expert and the ASEs, we utilize the complete training and development partitions of the ASVspoof 2019 LA (19LA) database~\cite{wang2020asvspoof}. 
The shared expert \(E_0\) is trained on the original, unmodified audio data to learn generalizable features. 
Each ASE is trained with data subjected to a specific post-processing transformation designed to simulate particular attack scenarios. Specifically, we employ RawBoost \cite{tak2022rawboost} configurations 1, 2, and 3 to simulate different post-processing attacks and use them to train expert models \(E_1\) to \(E_3\). These transformations include linear and non-linear convolutive distortions (e.g., filtering and reverberation), impulsive signal-dependent additive noise and stationary signal-independent additive noise.

For other noise based attack simulation, we apply Gaussian random noise 
augmentations 
to train expert \(E_4\). Additionally, to simulate filter based attacks, we apply lowpass, bandpass and highpass filters
to generate training data for expert \(E_5\).


For training the AEF module, we randomly sample $25\%$ of the data used for training every experts, which corresponds to $6345$ samples per expert. 
Given our $6$ experts $\{E_i\}, i \in \{0, 1, \dots, 5\}$, we end up with $6 \times 6345 $ training samples.  


We evaluate our method on both clean and attacked versions of the 19LA evaluation set. 
Specifically, T0 denotes the original 19LA set, while T1–T3 are its variants attacked using the same RawBoost strategies applied during the training of experts $E_1, E_2$ and $E_3$, respectively. T4 and T5 include attacks used in training $E_4$ and $E_5$, along with additional unseen perturbations: T4 introduces unseen color noise, and T5 includes unseen filtering operations. 
Moreover, we evaluate our methodology also on the ASVspoof 2021 LA (21LA) database~\cite{yamagishi2021asvspoof}, which we define as T6. This set introduces unseen codec and transmission variability, which is crucial for assessing models' robustness. 
All data processing scripts and code will be publicly available at: \url{https://github.com/foolishcqx/AMULET-AudioSpoof}.


\begin{table}[t]
\centering
\caption{EER (\%) achieved by AMULET: FFT vs. LoRA. }
\vspace{-5pt}
\label{tab: performance compare}
\footnotesize
\begin{tabular}{lcc} 
\toprule
Model             & Avg. (T0-T5)     & T6  
\\ \midrule
\( E_0 \) + top-3 (FFT)   & \textbf{0.54~} & 1.41~           \\
\( E_0 \) + top-3 (LoRA) & 0.55~          & \textbf{1.19~}  \\ \midrule
\( E_0 \) + top-4 (FFT)   & \textbf{0.46~} & 1.46~           \\
\( E_0 \) + top-4 (LoRA) & 0.55~          & \textbf{1.17~}  \\ \midrule
\( E_0 \) + top-5 (FFT)   & \textbf{0.39~} & 1.36~           \\
\( E_0 \) + top-5 (LoRA) & 0.46~          & \textbf{1.10~}  \\\bottomrule
\end{tabular}
\vspace{-3pt}
\end{table}

\begin{table*}[t]
\centering
\caption{Performance, in terms of EER, of each expert, expert aggregation strategies and our method in mixture attack.}
\vspace{-5pt}
\label{tab: Miture attack}
\footnotesize
\begin{tabular}{lcccccccc} 
\toprule
Model          & NoiseFirst     & FilterFirst            & RawBoost 4     & RawBoost 5     & RawBoost 6     & RawBoost 7     & RawBoost 8     & Avg.             \\ \midrule
\( E_0 \) & 18.51~         & 18.29~                 & 26.24~         & 21.28~         & 6.80~          & 22.69~         & 8.91~          & 17.53~          \\
\( E_1 \)& 12.85~         & 11.86~                 & \underline{3.37~}  & 2.58~          & \underline{2.27}~  & 1.85~          & 0.57~          & 5.05~           \\
\( E_2 \)& 14.19~         & 13.84~                 & 4.83~          & 5.16~          & 3.00~          & \textbf{0.75}~ & 0.39~          & 6.02~           \\
\( E_3 \)& 10.94~         & 10.67~                 & 21.19~         & 17.47~         & 3.85~          & 14.23~         & 5.96~          & 12.04~          \\
\( E_4 \)& 10.75~         & 9.38~                  & 23.92~         & 24.43~         & 10.14~         & 18.09~         & 9.52~          & 15.18~          \\
\( E_5 \)& \underline{6.62~}  & \textbf{5.97~}         & 15.83~         & 10.32~         & \textbf{2.16~} & 12.80~         & 3.13~          & 8.12~           \\ \midrule
\textit{Ensemble}       & 10.12~         & 9.44~                  & 4.35~          & 1.86~          & 3.50~          & 2.22~          & 0.46~          & 4.56~           \\ 
\textit{Attention} \cite{zhou2022adaptive} & 14.37~         & 15.20~                 & 12.77~         & 4.16~          & 7.91~          & 8.12~          & 0.75~          & 9.04~           \\
\textit{MoEF} \cite{wang2025mixture}& 7.45~          & 7.06~                  & 4.82~          & 4.00~          & 2.94~          & 1.63~          & 1.32~          & 4.18~           \\ \midrule
\( E_0 \) + top-3 (Ours) & 7.07~          & 6.54~                  & 3.68~          & \underline{1.74}~  & 2.73~          & 1.58~          & \textbf{0.31}~ & 3.38~           \\
\( E_0 \) + top-4 (Ours) & 7.39~          & 6.69~                  & 3.51~          & 1.76~          & 2.55~          & 1.28~          & \underline{0.34}~  & \underline{3.36}~   \\
\( E_0 \) + top-5 (Ours) & \textbf{6.39}~ & \underline{6.05~} & \textbf{3.26~} & \textbf{1.66}~ & 2.51~          & \underline{1.16}~ & \textbf{0.31~} & \textbf{3.05~}  \\
\bottomrule
\end{tabular}
\vspace{-10pt}
\end{table*}

\subsection{Implementation Details} \label{Data Augmentation}

During LoRA fine-tuning, all layers within the W2V2 model are updated using low-rank decomposition with rank $r = 8$, scaling factor $\alpha = 32$ and dropout rate of $0.1$. The bias terms remain fixed during this process. Fine-tuning is performed with a learning rate of $10^{-4}$. 
We reduce the learning rate by $0.5$ after three consecutive epochs without improvement, with a lower bound set at $10^{-7}$, and we use early stopping with patience $10$.
The model achieving the lowest development Equal Error Rate (EER) is selected for evaluation. 
All other hyperparameters align with those specified in \cite{tak2022automatic}.
The best results are highlighted in bold, while the second-best results are underlined. 

\vspace{-5pt}
\subsection{Baselines} 
\label{baselines}
In addition to our expert-based models ($E_0$–$E_5$), several baseline methods are incorporated for comparative analysis. The \textit{Ensemble} baseline computes the mean of the logit outputs from the LoRA-finetuned ASEs and the shared expert without requiring additional training. The \textit{Attention} and \textit{MoEF} baselines implement expert aggregation strategies proposed in \cite{zhou2022adaptive} and \cite{wang2025mixture}, respectively, replacing the AEF module in AMULET while keeping the remaining training procedures identical. 

\vspace{-5pt}

\subsection{Performance Under Single Post-Processing Attack}

In Table~\ref{tab:Mode Performance}, we present the performance of each expert model, the baseline methods, and our proposed approach under a single post-processing attack setting (i.e., across the previously introduced T0–T6 datasets). In specific, the configuration denoted as ``$E_0$ + top-$k$'' represents our adaptive expert fusion strategy, in which a gating network dynamically selects the top $k$ ASEs to be activated together with the base expert $E_0$. 

As expected, across the ASEs $\{E_i\}, i \in \{ 1, 2, \ldots, 5 \}$, every expert achieves the best results on its matching test set. For instance, $E_1$ achieves the best EER on T1. 
Our method achieves performance comparable to individual expert models while maintaining robustness across datasets with varying expert performance. 
On average, we also outperform the other baselines. This improvement can be attributed to the AEF module, which dynamically assigns expert weights based on input characteristics, effectively integrating complementary attack-specific knowledge. By selectively combining experts with specialized knowledge, we adapt to different spoofing and noise conditions, thus improving performance even under challenging attack scenarios. 

\vspace{-5pt}
\subsection{Efficiency Analysis}
We now perform the comparison between the performances achieved by the various ASEs in case of fine-tuning with LoRA (our proposed approach) vs the full fine-tuning (FFT). 
As shown in Table~\ref{tab:experts performance}, three experts out of five performs even better with the LoRA fine-tuning. This is noteworthy, considering that the LoRA strategy allows to use only $1.13\%$ of the trainable parameters, significantly reducing computational costs without sacrificing accuracy. 

Considering the entire AMULET framework, in Table~\ref{tab: performance compare} we show that, while FFT provides better robustness against known post-processing operations (i.e., on T1-T5 datasets), LoRA fine-tuning achieves nearly identical performance. 
On datasets with unknown post-processing methods (i.e., on T6 dataset), LoRA fine-tuning results in a lower EER, indicating better robustness. This highlights AMULET’s efficiency in maintaining robustness while minimizing computational overhead, making it a scalable solution for real-world applications.

\vspace{-5pt}
\subsection{Performance Under Mixed Post-Processing Attacks}
Building on previous experiments, we further evaluate the AMULET framework’s robustness against mixed post-processing attacks, where multiple techniques are combined to simulate real-world adversarial scenarios. To this end, we apply hybrid RawBoost transformations alongside noise and filtering effects
to the 19LA dataset. The RawBoost transformations involve the combined application of RawBoost 1, 2, and 3.
Additionally, we define two conditions, NoiseFirst and FilterFirst, to specify the sequence in which noise and filtering are applied.

As shown in Table \ref{tab: Miture attack}, single expert models exhibit significant performance variations under mixed attacks, indicating their limited robustness. Although other baseline strategies such as \textit{Ensemble} and \textit{MoEF} partially address this issue, our framework consistently delivers superior and stable performance across various mixed post-processing attacks. 
This demonstrates AMULET's effectiveness and robustness across multiple datasets, highlighting its ability to handle diverse adversarial conditions and reinforcing its suitability for real-world deployment.

\section{Conclusion}
This paper proposed the AMULET framework for robust audio spoofing detection, leveraging Attack-Specific Experts (ASEs) and Adaptive Expert Fusion (AEF) to adapt to diverse post-processing and noisy conditions. Experiments demonstrate that AMULET achieves high accuracy across various settings, with LoRA fine-tuning offering comparable performance to full fine-tuning at reduced cost. AMULET also outperforms single experts and other fusion methods, especially in mixed or unpredictable attacks. 

\bibliographystyle{IEEEtran}
\bibliography{reference}

\end{document}